\documentclass[submission,copyright,creativecommons]{eptcs}
\providecommand{\event}{P\&C 2018} 
\usepackage{breakurl}             
\usepackage{physics}
\usepackage{qcircuit}
\usepackage{graphicx}
\usepackage{caption}
\usepackage{subcaption}
\usepackage{amsmath}
\usepackage{dsfont}
\title{One-Sided Device-Independent Certification of Unbounded Random Numbers}
\author{Brian Coyle
\institute{School of Informatics, University of Edinburgh,}
\institute{Edinburgh EH8 9AB, United Kingdom.}
\email{brian.coyle@ed.ac.uk}
\and
Matty J. Hoban 
\institute{Department of Computer Science, University of Oxford,}
\institute{Oxford OX1 3QD, United Kingdom.}
\email{matthew.hoban@cs.ox.ac.uk}
\and
Elham Kashefi
\institute{School of Informatics, University of Edinburgh,}
\institute{Edinburgh EH8 9AB, United Kingdom.}
\institute{Laboratoire  d’Informatique  de  Paris  6,  CNRS,}
\institute{Sorbonne  Universit́e,  4  place  Jussieu,  75005  Paris.}
\email{ekashefi@inf.ed.ac.uk}
}
\def\titlerunning{One-Sided Device-Independent Certification of Unbounded Random Numbers}
\def\authorrunning{B. Coyle, M. J. Hoban \& E. Kashefi}

\begin{document}
\maketitle

\begin{abstract}
The intrinsic non-locality of correlations in Quantum Mechanics allow us to certify the behaviour of a quantum mechanism in a device independent way. In particular, we present a new protocol that allows an unbounded amount of randomness to be certified as being legitimately the consequence of a measurement on a quantum state. By using a sequence of non-projective measurements on single state, we show a more robust method to certify unbounded randomness than the protocol of \cite{curchod_unbounded_2017}, by moving to a \textit{one-sided} device independent scenario. This protocol also does not assume any specific behaviour of the adversary trying to fool the participants in the protocol, which is an advantage over previous steering based protocols. We present numerical results which confirm the optimal functioning of this protocol in the ideal case. Furthermore, we also study an experimental scenario to determine the feasibility of the protocol in a realistic implementation. The effect of depolarizing noise is examined, by studying a potential state produced by a networked system of ion traps.
\end{abstract}

\section{\label{sec:intro}Introduction}

Quantum mechanics is a theory that can exhibit, in some sense, fundamental randomness. This randomness can be extracted by measurements on a quantum system, but if the party preparing the quantum state and/or measurement apparatus is untrusted, how can we verify that a true measurement is occurring on a real quantum state? The seed of the answer was discovered by Bell in \cite{bell_einstein_1964} in the form of Bell inequalities. The violation of these inequalities by certain quantum systems proved, along with the argument of Einstein, Podolsky and Rosen in 1935, \cite{einstein_can_1935}, that quantum mechanics must be a non-local theory. Using these inequalities and non-local properties, it is possible to test the following about a process. If we acquire statistics produced by some procedure, and it can be shown that the statistics violate these Bell inequalities, then those statistics cannot have been produced by a local hidden variable theory. Using these ideas, it is possible to determine if randomness produced by a given apparatus was in fact the result of measurements on a quantum system, as opposed to being the result of a deterministic process. This `Bell non-locality' has been utilized extensively in a situation referred to as \textit{device independence}, where it is possible to certify quantum behaviours, even in a scenario where the party producing the device is untrusted, \cite{pironio_random_2010, colbeck_quantum_2009}. \\
\indent In \cite{curchod_unbounded_2017}, the authors propose a scenario to generate and certify an unbounded amount of randomness using a sequence of non-projective measurements on a single quantum state. Non-projectivity is required to preserve some entanglement in the quantum state after the measurement, which is an essential resource in determining non-locality. In this scenario, Alice (A), and Bob (B), share an entangled state, possibly produced by a third party eavesdropper, Eve (E). Both halves of this state are contained in two separate devices and each sent to one of Alice or Bob. Alice (Bob) then chooses to measure in a particular basis, $x$ ($y$), and record their measurement outcome, $a$ ($b$). The randomness of one (local randomness), or both (global randomness) outcomes can be certified using a violation of a particular Bell inequality. Both inputs, $x,y$ and outcomes $a,b$ are assumed to be binary variables for simplicity, such that:
\begin{align*}
\overbrace{
	\begin{array}{ll}
		x  \in \left\{0,1\right\} \\
		y  \in \left\{0,1\right\} \\
	\end{array}
	}^\text{Measurements:} \qquad 
\overbrace{
	\begin{array}{ll}
		a  \in \left\{1,-1\right\} \\
		b  \in \left\{1,-1\right\} \\
	\end{array}
}^\text{Outcomes:}
\end{align*} 
\begin{figure}[h!]
        \centering
            \includegraphics[width = \linewidth, height =0.3\linewidth]{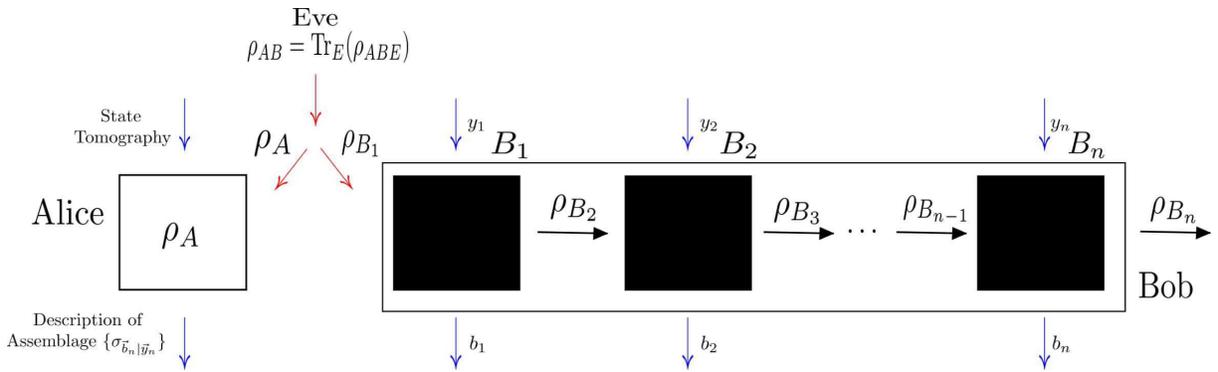}
            \caption{Illustration of Protocol 1. Bob makes a sequence of measurements on his state in a black box, and Alice certifies the randomness of the outcomes using the assemblage, $\{\sigma_{\vec{b}_n|\vec{y}_n}$\}}
            \label{fig:seriessteering}
\end{figure}

It is also possible to lift the trust restrictions on one of the parties, Alice say, so that she has full autonomy over her half of the shared state and measurement apparatus. This means that at any stage in the process, she has the ability to do quantum state tomography to determine the state she possesses and she is able to directly control her measurements. This is referred to as the \textit{steering} or \textit{one-sided device independent} scenario, because the results of Bob's measurements on his side of the shared state cause Alice to be `\textit{steered}' into a certain state, which is dependent on Bob's measurement choice. Typically, in the fully device independent scenario, the joint probability distribution between Alice and Bob's outcomes given their measurement choices, $P(ab|xy)$, is the relevant quantity studied to enable certification of randomness. However, in the one-sided case, this is no longer relevant because we are not interested in Alice's measurement outcomes. Instead, we study the \textit{assemblage}, $\{\sigma_{b|y}\}$, \cite{skrzypczyk_steeringreview:_2016}, which is the set of conditional unnormalized quantum states that one party can be steered into, given measurement choices of the other. In this paper, we will keep the convention of \cite{curchod_unbounded_2017}, where Bob makes measurements on his state, and Alice's state is the one which is steered. These assemblage elements, $\sigma_{b|y}$, are conditional on Bob's outcome, $b$, and his choice of measurement basis, $y$. The elements are defined by: $\sigma_{b|y} = p(b|y)\rho_{b|y}=  \tr_B[(\mathds{1}_A \otimes M_{b|y})\rho_{AB}]$, where $\rho_{AB} = \tr_E{\rho_{ABE}}$ is the state prepared by Eve and sent to Alice and Bob, and $\rho_{b|y}$ is the state that Alice is steered into conditional on Bob's input, $y$, and measurement outcome, $b$. $M_{b|y}$ are the POVM elements Bob expects to be able to measure by choosing his input $y$, such that $M_{b|y} \geq 0 \ \forall b,y$, and $\sum_b M_{b|y} = \mathds{1}, \forall y$.\\
\indent In the steering scenario, the certification of the local randomness of Bob's outcomes can be done by examining the assemblage elements, and their violation of `\textit{steering inequalities}', analogous to the violation of Bell inequalities by the non-local probability distributions, $P(ab|xy)$. A maximal violation of steering inequalities corresponds to a maximally steerable state. The idea of producing certifiable randomness using steering was first studied by Law \textit{et al.} \cite{law_quantum_2014}, and then with the assistance of semi-definite programming by Passaro \textit{et al.} \cite{passaro_optimal_2015}. Given an assemblage, a method was derived to determine the \textit{steerability} of the assemblage via semi-definite programs (SDPs) by Skrzypczyk \textit{et. al.}, \cite{skrzypczyk_quantifying_2014}. The \textit{steering weight} (SW) is given to be the solution to the following SDP, (\ref{steeringweightsdp}), and its dual program, (\ref{steeringweightdual}):
\begin{tabular}{p{5cm} p{2cm} p{5cm}}
\begin{align}
       { \begin{array}{lllll|}
        SW=& \min        & 1- \tr\sum\limits_\lambda\sigma_\lambda &\\ 
           & \text{s.t. }&\sigma_{b|y} - \sum\limits_{\lambda}D(b|y,\lambda)\sigma_\lambda \geq 0 & \forall b,y\\ 
           &      & \sigma_{\lambda} \geq 0,    &\forall \lambda 
        \end{array} \label{steeringweightsdp}}
    \end{align} & &\begin{align}
       { \begin{array}{|lllll}
            SW =& \max        & 1 - \tr\sum\limits_{by}F_{b|y}\sigma_{b|y}                &\\ 
                &\text{s.t. } &  \sum\limits_{by}D(b|y,\lambda)F_{b|y}-\mathds{1} \geq 0  &\forall \lambda\\ 
                &             & F_{b|y}  \geq 0,                                          & \forall b,y
        \end{array} \label{steeringweightdual}}
    \end{align}\\
\end{tabular}

The dual program, (\ref{steeringweightdual}), is the most relevant for this paper because, as shown in \cite{skrzypczyk_quantifying_2014}, the dual variables of the SDP, (\ref{steeringweightdual}), in fact define a steering inequality, $\{F_{b|y}\}$, for which the assemblage, $\{\sigma_{b|y}\}$, produces a maximal violation. $\{\sigma_\lambda\}$ is an assemblage that Eve could produce for Alice using hidden variables, $\lambda$, and the SDPs, (\ref{steeringweightsdp}), (\ref{steeringweightdual}), test for the existence of such an assemblage. In the case where certifiable randomness is produced as a result of Bob's measurement, we want no such \textit{local hidden state} (LHS) assemblages to exist. If Eve had the ability to reproduce the assemblage that Alice receives, by using her knowledge of these hidden variables then, from Eve's point of view, the outcomes that Bob receives are in fact deterministic, and not random. \\
\indent The scenario that this work is presented in is similar to that of \cite{curchod_unbounded_2017}, where we assume Bob can implement non-projective measurements in rotated versions of the Pauli-$X$ and $Z$ bases, however Alice only needs the functionality to implement projective Pauli-$X$ and $Z$ basis measurements, since it is sufficient for her to do quantum state tomography to certify Bob's random outcomes. Also, \cite{curchod_unbounded_2017} only considers the X basis measurement to be non-projective, and hence the random outcomes are obtained from a sequence of these X measurements. However, in this case, it is possible for Bob to also to choose to measure in a non-projective Z basis also. The motivation for this is the following. If Bob has the following state: $\ket{0}$, and makes a measurement in the Pauli-X basis, $\ket{\pm} = \frac{1}{\sqrt{2}}(\ket{0}+\ket{1})$, he will get one of the outcomes $b_1 = \pm 1$, each with probability $1/2$. If he then makes a second measurement on the state, this time in the Pauli-$Z$ basis, he will get one of the outcomes $b_2 = \pm 1$ each with probability $1/2$. However, if he had chosen his second measurement to be in the X basis, he would not get a random result, but a deterministic one. This example illustrates that measuring in (almost) orthogonal bases should give the maximal amount of randomness and will be further reinforced by numerical evidence shown in Figures~(\ref{fig:TwoRounds2}) and (\ref{fig:ThreeRounds2}) in Section~\ref{ssec:ideal} .\\
\indent The following non-projective Kraus operators, $\Pi^{x(\theta)}_{b|y}$ are defined in \cite{curchod_unbounded_2017}:
\begin{align}
\Pi^{x(\theta)}_{\pm 1|1} &= \cos(\theta)\ket{\pm}\bra{\pm}+\sin(\theta)\ket{\mp}\bra{\mp}\label{fdikrausoperators} 
\end{align}
These Kraus operators, which will be denoted by a measurement in the $X_\theta$ basis, reduce to the usual Pauli-X basis measurement operators for $\theta = 0$.\\ 
\indent Introducing non-projective Z basis measurements corresponds to defining the following operators, denoted by $Z_\phi$, which again reduce to the usual computational basis measurements for $\phi = 0$. The Kraus operators for these non-projective measurements are given by:

\begin{align}
\Pi^{z(\phi)}_{1|0} &= \cos(\phi)\ket{0}\bra{0}+\sin(\phi)\ket{1}\bra{1}\label{1sdikrausoperators} \qquad
\Pi^{z(\phi)}_{-1|0} =  \cos(\phi)\ket{1}\bra{1}+\sin(\phi)\ket{0}\bra{0}
\end{align}
Therefore, the POVM that Bob implements on his half of the shared state is:
\begin{align}
M^{x(\theta)/z(\phi)}_{b|y} = (\Pi^{x(\theta)/z(\phi)}_{b|y})^\dagger(\Pi^{x(\theta)/z(\phi)}_{b|y})
\end{align}

with $y = 1$ indicating that he has chosen to measure in the non-projective $X_\theta$ basis, and $y = 0$ indicates a measurement in the $Z_\phi$ basis. For a single measurement, if Alice wants to certify the randomness produced by Bob's non-projective $X_\theta$ measurements, the protocol should be repeated, with the same state produced by Eve, but in this `test' run, Bob will choose to measure in the $Z_0$ basis. Alice will then do state tomography on the resulting states to determine them, and repeats until she has gathered enough statistics to reproduce the full assemblage with high enough confidence. \\
\indent The quantifier of certifiable randomness that will be used is the \textit{guessing probability} (GP), $P_{\text{G}}$. This quantifier was first discussed in \cite{pironio_random_2010} and was used in \cite{curchod_unbounded_2017} for fully device-independent randomness certification. The SDP used in Protocol 1 is similar to that of \cite{passaro_optimal_2015}, where the authors define the guessing probability in terms of the local hidden state (LHS) strategies that Eve could use to produce set of states for Alice and Bob which are determined by local hidden variables known only to her. However, this method uses an assumption about the fact that Eve creates these assemblage elements using local measurements on \textit{her} side of the entangled state, effectively steering Alice and Bob into a given state, about which she could deduce certain properties. The ability for Eve to do this is clearly undesirable as this would enable her to have extra information about Bob's random outcomes. Essentially, this means his outcomes would be reproducible by some local hidden state model that Eve is using, as described above. However, the results of this paper make no assumptions about the specific actions of Eve. For clarity, we will study the case of a single measurement before giving the results for a sequence of measurements. With just a single measurement, the GP is given as the solution to the following SDP:

\begin{align}
    \begin{array}{ccll}
        P_{\text{G}}(y=y^*) =& \max\limits_{\{\sigma^{E}_{b|y}\}_{b,y}}&\tr_A[\sigma^E_{b|y^*}] &\\ 
        &\text{s.t. } &\sum\limits_{b,y}F_{b|y}\sigma_{b|y}^E = v& \label{gpsdprelax}\\ 
        &&\sum\limits_{b}\sigma_{b|y}^E = \sum\limits_{b}\sigma_{b|y'}^E & \forall e, y \neq y'\\ 
        &&\sigma_{b|y}^E\succeq 0 & \forall y,b
    \end{array}
\end{align}

The steering inequality $\{F_{b|y}\}$ is the one determined by the SDP, (\ref{steeringweightdual}), which is maximally violated by the ideal assemblage, $\{\sigma_{b|y}\}$, that Alice expects to have access to if Eve follows the protocol honestly. The SDP, (\ref{gpsdprelax}), allows Eve to create, for Alice, any assemblage, $\{\sigma_{b|y}^E\}$, as long as this assemblage obeys the constraints in the SDP. The first constraint enforces the fact that this assemblage should produce a violation of the steering inequality, $\{F_{b|y}\}$, with violation $v$ that would be produced by the ideal assemblage. The second constraint enforces that Alice and Bob cannot communicate faster than the speed of light (no-signalling condition), while the last constraint enforces that Eve must produce a valid assemblage for Alice i.e. it must be a positive semidefinite matrix. We also assume Eve knows the measurement setting from which Bob wants to extract randomness, $y = y^*$.\\
\indent Once Bob has made his measurement, Alice can then determine the state she then possesses as a result. By repeating multiple runs of the protocol, Alice can determine the full statistics of Bob's measurement outcomes and hence the full assemblage. Once she knows the assemblage produced by the given initial state and measurement set, she can then calculate the optimal steering inequality for that assemblage, using (\ref{steeringweightdual}), and the associated value of the violation, $v$, given by the steering inequality. Using this, she can calculate the GP with the SDP, (\ref{gpsdprelax}). This guessing probability, as discussed in (\cite{curchod_unbounded_2017}, \cite{pironio_random_2010} , \cite{law_quantum_2014}, \cite{passaro_optimal_2015}) is the optimal probability that Eve can guess Bob's outcome, $b$, given any information that she possesses. For example, if the assemblage is unsteerable (it has a steering weight of 0), then it is unsteerable with respect to any steering inequality and so the value of the violation, $v$, will reflect this. In this scenario, Eve could have engineered Bob's device to include some local hidden variables and hence produce deterministic outcomes. It is exactly this situation which we want to detect. For a single measurement, if the GP is equal to $1/2$, the outcome of the measurement is in fact random and Eve's only strategy is simply to guess randomly which outcome Bob received. However, if it is equal to 1, Eve knows the outcome exactly since, from her point of view, the process was deterministic. Clearly, to optimally certify randomness, we want the GP to be as close to the former situation as possible.\\
\indent A further quantifier which is useful is the \textit{min entropy}, $H_{min}$, \cite{curchod_unbounded_2017}:

\begin{align}
    H_{min} = -\log_2(P_G) \label{minentropy}
\end{align}
The meaning of this quantity is clear. If $P_G = 1/2 \implies H_{min} = 1$ and so one certifiable random bit is produced by the measurement. If $P_G = 1 \implies H_{min}= 0$ and no randomness can be certified, i.e. the assemblage could have been produced by a LHS model.\\

\section{One-Sided Device-Independent (1SDI) Protocol}\label{sec:protocol}

As in \cite{curchod_unbounded_2017}, we can extend this scenario to one in which Bob implements a sequence of non-projective measurements on his half of the shared state. Defining the protocol for $n$ rounds is therefore straightforward ($n$ is predetermined by Alice and Bob), where on each round, Bob makes one measurement on the shared state. Bob will input his choice of measurement basis for the $n$ rounds, denoted $\vec{y}_n = y_1y_2...y_n \in \{0,1\}^n$, into the device and record his measurement outcomes, denoted $\vec{b}_n = b_1b_2...b_n \in \{1,-1\}^n$. In round $k$, Bob chooses to measure in the `\textit{noisy}' Pauli-X basis, $X_{\theta_k}$, or the `\textit{noisy}' Pauli-Z basis, $Z_{\phi_k}$, using the Kraus operators defined by (\ref{fdikrausoperators}), (\ref{1sdikrausoperators}) respectively. Of course, since the scenario is device independent, Bob does not know if these measurements were actually performed in the device, until the randomness is finally certified by the protocol. \\
\indent If Alice wants to certify the randomness of the outcomes for all rounds up to round $n$, she must find the solution for the SDP, (\ref{1sdisdpsteeringinequality}), for all $k < n$. This set of SDP's will give her the optimal steering inequality for each round $k$ ($1 \leq k \leq n$), $\{F_{\vec{b}_k|\vec{y}_k}\}$, which is maximally violated by the assemblage $\{\sigma_{\vec{b}_k|\vec{y}_k}\}$. 
\begin{align}
    \begin{array}{ccll}
        SW(\sigma_{\vec{b}_k|\vec{y}_k}) =& \max& 1 - \tr\sum\limits_{\vec{b}_k,\vec{y}_k}F_{\vec{b}_k|\vec{y}_k}\sigma_{\vec{b}_k|\vec{y}_k}&\\ 
        &\text{s.t. } &  \sum\limits_{\vec{b}_k,\vec{y}_k}D(\vec{b}_k|\vec{y}_k,\vec{\lambda}_k)F_{\vec{b}_k|\vec{y}_k}-\mathds{1} \geq 0 &\forall \vec{\lambda}_k\\ 
        && F_{\vec{b}_k|\vec{y}_k}  \geq 0, & \forall \vec{b}_k, \vec{y}_k
    \end{array} \label{1sdisdpsteeringinequality}
\end{align}

This SDP calculates the steering weight for the assemblage created on measurement round $k$, however the actual value of this steering weight is not important for our purposes. Instead, we want to extract the dual variables, $\{F_{\vec{b}_k|\vec{y}_k}\}$, which again define a steering inequality.\\
\indent This SDP is adapted from \cite{skrzypczyk_quantifying_2014} and as in that case, the primal SDP checks a given assemblage against all possible deterministic strategies, $D(\vec{b}_k|\vec{y}_k,\vec{\lambda_k})$. This determines if the assemblage can be decomposed as a convex combination of assemblages, $\sigma_{\vec{\lambda}_k}$ that Eve could have created in some LHS model, given her possible knowledge of $k$ hidden variables, $\vec{\lambda}_k = \lambda_1\lambda_2...\lambda_k$. Again, the steering inequality can be decomposed into a linear combination of these assemblage elements, with coefficients given by the variables $F_{\vec{b}_k|\vec{y}_k}$, which are the dual variables in the SDP, (\ref{1sdisdpsteeringinequality}). Once Alice has this set of steering inequalities, she can determine the guessing probability for Eve, as the solution of the following SDP, (\ref{mainsdp}):
\begin{small}
\begin{align}
&P_{\text{G}}(\vec{y}_n^*, F_{\vec{b}_n|\vec{y}_n}) = \max\limits_{\vec{b}_n,\vec{y}_n} \tr_A\sigma^E_{\vec{b}_n|\vec{y}_n = \vec{y}_n^*}\label{mainsdp}\\
 \text{s.t. } &\begin{array}{|c|c|}
\hline
\begin{array}{lll}
& \sum\limits_{\vec{b}_n,\vec{y}_n} F_{\vec{b}_n|\vec{y}_n}\sigma^E_{\vec{b}_n|\vec{y}_n} = v_n, &  \\
& \sum\limits_{\vec{b}_{n-1},\vec{y}_{n-1}} F_{\vec{b}_{n-1}|\vec{y}_{n-1}}\sigma^E_{\vec{b}_{n-1}|\vec{y}_{n-1}} = v_{n-1} & \\ \nonumber
&\vdots& \vdots\\
& \sum\limits_{b_1,y_1} F_{b_1|y_1}\sigma^E_{b_1|y_1} = v_1, & \\
\end{array} & \begin{array}{lll}
& \sum\limits_{b_n} \sigma^E_{\vec{b}_n|\vec{y}_n} = \sigma^E_{\vec{b}_{n-1}|\vec{y}_{n-1}} ,& \qquad \forall y_n \\
& \sum\limits_{b_{n-1}} \sigma^E_{\vec{b}_{n-1}|\vec{y}_{n-1}} = \sigma^E_{\vec{b}_{n-2}|\vec{y}_{n-2}} ,& \qquad \forall y_{n-1} \\
&\vdots &\qquad \vdots\\
& \sum\limits_{b_1}\sigma^E_{b_1|y_1} = \rho_A & \qquad \forall y_1 
\end{array}\\ \hline
\begin{array}{lll}
&\sum\limits_{\vec{b}_n} \sigma^E_{\vec{b}_n|\vec{y}_n} = \sum\limits_{b_n} \sigma^E_{\vec{b}_n|\vec{y}_n'}, & \forall \vec{y}_n,\vec{y}_n' \\
& \sum\limits_{\vec{b}_{n-1}} \sigma^E_{\vec{b}_{n-1}|\vec{y}_{n-1}} = \sum\limits_{b_{n-1}} \sigma^E_{\vec{b}_{n-1}|\vec{y}_{n-1}'}, & \forall \vec{y}_{n-1},\vec{y}_{n-1}' \\
&\vdots & \vdots\\
& \sum\limits_{b_1}\sigma^E_{b_1|y_1} = \sum\limits_{b_1}\sigma^E_{b_1|y_1'}, & \forall y_1,y_1' \\
\end{array} & \begin{array}{llll}
& \sigma^E_{\vec{b}_n|\vec{y}_n} \succeq 0, &  \qquad  \forall \vec{y}_n,\vec{b}_n \\
& \sigma^E_{\vec{b}_{n-1}|\vec{y}_{n-1}} \succeq 0, & \qquad \forall \vec{y}_{n-1},\vec{b}_{n-1} \\
&\vdots & \qquad \vdots&\\
& \sigma^E_{b_1|y_1} \succeq 0 & \qquad \forall y_1, b_1 \\
\end{array}\\\hline
\end{array}\nonumber 
\end{align}
\end{small}

Where the solution of this SDP is the guessing probability and the maximum over the trace of all the assemblages that Eve can create for Alice at the end of the protocol, $\sigma^E_{\vec{b}_n|\vec{y}_n = \vec{y}_n^*}$, for a particular input string, $\vec{y}^*_n$. Again, Eve knows from which measurement settings, $\vec{y}_n^*$, Bob wants to extract randomness. The steering inequality violations, $\vec{v}_n = v_1v_2\dots v_n$ can be calculated by Alice once she has determined the associated steering inequality (if one exists). The constraints of the SDP are similar to the single measurement case except for the addition of one new set of constraints which are required for a sequence. These particular constraints enforce causality in the measurement sequence, so that, for example (for two measurement rounds): 
\begin{align}
\sum\limits_{b_{2}} \sigma^E_{b_1b_2|y_1y_2} = \sigma^E_{b_{1}|y_{1}}, \qquad \forall y_2
\end{align}
Simply put, this constraint means that Eve has no access to future events, i.e. in measurement round $i$, she only has access to information from rounds $j < i$ to aid in her attempts to guess the measurement outcomes.\\
\indent For the final measurement round, the measurement operators become projective to end the protocol, i.e. $\theta_n = \phi_n = 0$ and the state at round $n-1$ is a pure entangled state. In this case, it is possible to define the steering inequality explicitly, as done in \cite{skrzypczyk_quantifying_2014}:
\begin{align}
F_{\vec{b}_n|\vec{y}_n} &= \alpha\left(\mathds{1} - \frac{\sigma_{\vec{b}_n|\vec{y}_n}}{\tr(\sigma_{\vec{b}_n|\vec{y}_n})}\right) \label{projectivesteeringfunctional}
\end{align}
where $\alpha$ is chosen sufficiently large. A choice of $\alpha = 100$ was chosen for all numerical results in this paper. Clearly, this choice of a steering inequality automatically gives a maximal violation value of $v_n = 0$. \\
\indent Protocol 1 describes the full scenario in detail. If the guessing probability after $n$ measurement rounds is sufficiently close to $1/2^n$, then Eve has followed the protocol faithfully and produced the required quantum state and measurement apparatus for Bob. This means that the probability of Eve guessing the sequence of bits that Bob has obtained decreases exponentially with the number of rounds in the protocol and we have true quantum randomness.\\

\noindent\textbf{Protocol 1}: \underline{1SDI Randomness Certification}
\begin{enumerate}
    \item Eve prepares joint state $\rho_{ABE}$ \& sends state $\rho_{AB} = \tr_E(\rho_{ABE})$ to Alice and Bob. Bob's state is contained in a black box with the ability to implement a predetermined measurement sequence with angles, $\{\theta_1, \theta_2,\dots, \theta_n\}, \{\phi_1, \phi_2,\dots \phi_n\}.$
    \item Bob chooses measurement $y_1^*$ and makes measurement, $M_{b_1|y_1^*}$, on state corresponding to a measurement in either $X_{\theta_1} (y^*_1=1)$, or $Z_{\phi_1} (y^*_1=0)$ basis.
    \item Alice's state is steered into $\sigma_{b_1|y_1^*}$, which she determines using state tomography.
    \item Alice and Bob repeat step 2. and 3. up to $n$ rounds to determine full assemblage for each round, $k$, $\{\sigma_{\vec{b}_k|\vec{y}^*_k}\}$ until Bob has made sufficient measurements to determine the measurement statistics accurately enough.
    \item Alice determines the steering inequality for each assemblage generated by each measurement round, $k$, $\{F_{\vec{b}_k|\vec{y}_k^*}\}$ using SDP, (\ref{1sdisdpsteeringinequality}), and the associated value of the steering inequality violation, $v_k$.
    \item Alice uses SDP, (\ref{mainsdp}), to determine the guessing probability for the assemblage after $n$ rounds.
    \item If the GP is sufficiently high, Alice and Bob abort the protocol and discard the measurement outcomes.
\end{enumerate} 

\indent Figure~(\ref{fig:seriessteering}) illustrates the protocol by writing Bob as a series of `Bob's' to illustrate the causal structure of the protocol. In this picture, each $B_i$ makes a single measurement on the state he receives from $B_{i-1}$ by choosing a basis $y_i$, and receiving measurement outcome $b_i$ before `passing' the state onto $B_{i+1}$. As described above, each Bob has no access to the black box he receives, but Alice has full autonomy over her device.

\subsection{Quantum Circuit for Protocol 1}\label{s:circuits}

The the following circuit, (\ref{circuitclassicalcontrol}), was designed to implement Bob's half of the protocol, with his sequence of $n$ measurements on his half of the shared state. First of all, the following two qubit unitary gates need to be introduced, that effectively implement the non-projective $X_\theta$ and $Z_\phi$ measurements, (\ref{nonprojectivegates})  respectively.

\begin{small}
\begin{align}
\Qcircuit @C=0.7em @R=0.7em {
\lstick{\ket{0^i}} & \multigate{3}{X_\theta}& \qw &=& & \qw &\gate{R_{y}(2\theta)} & \gate{H} & \ctrl{3} & \gate{H}  & \qw &\qw\\
\lstick{} &\ghost{X_\theta}& \qw & & & \qw &\qw & \qw & \qw & \qw & \qw &\qw\\
\lstick{}  &\ghost{X_\theta}& \qw & & & \qw &\qw & \qw & \qw & \qw & \qw &\qw \inputgroupv{2}{3}{.8em}{.8em}{\ket{0^k}} \\
\lstick{\ket{\psi}_B}& \ghost{X_\theta} & \qw& &&\qw & \qw & \qw& \gate{X} &\qw &\qw &\qw \gategroup{1}{7}{4}{10}{2em}{--} 
} 
\qquad \qquad
\Qcircuit @C=0.7em @R=0.7em {
\lstick{\ket{0^i}} & \multigate{3}{Z_\phi}& \qw &=& & \qw &\gate{R_{y}(2\phi)} & \gate{H} & \ctrl{3} & \gate{H}  & \qw &\qw\\
\lstick{} &\ghost{Z_\phi}& \qw & & & \qw &\qw & \qw & \qw & \qw & \qw &\qw\\
\lstick{}  &\ghost{Z_\phi}& \qw & & & \qw &\qw & \qw & \qw & \qw & \qw &\qw \inputgroupv{2}{3}{.8em}{.8em}{\ket{0^k}} \\
\lstick{\ket{\psi}_B}& \ghost{Z_\phi} & \qw& &&\qw & \qw & \qw& \gate{Z} &\qw &\qw &\qw \gategroup{1}{7}{4}{10}{2em}{--} 
}\label{nonprojectivegates}
\end{align}
\end{small}
In each gate above, the control ancilla is the topmost one, labelled by $i$, where all the other ($k$) ancillas pass through the gate acted on by the identity. The index on the ancilla will represent the measurement round it is used in. The input string, $\vec{y}_n$, for $n$ measurement rounds is used as classical input to the circuit, and conditioned on this input for each round, either the noisy X or noisy Z measurement is implemented. As mentioned above, it is the topmost ancilla that is used as a control qubit for each gate in the circuit. At the end of the protocol, all ancillas can be measured in the usual computational basis, where $Z^{(k)}$ represents the measurement done in round $k$. Clearly, if the input $y_k = 0$, the noisy Z measurement is implemented, $Z_{\phi_k}^{y_k\oplus 1}$, while if $y_k = 1$, the noisy X measurement is implemented, $X_{\theta_k}^{y_k}$, and the other is not. In this fashion, only one quantum gate acts on the state per measurement round. Also, the state $\rho_B$ is only Bob's half of the initial state.\\
\indent This circuit is designed in the standard manner, in which all measurements are deferred to the end of the circuit. However, it could be further improved by using only a single ancilla. This ancilla would undergo multiple measurements, with the addition of a series of CNOT gates to the ancilla wire. These CNOT gates would return the ancilla to the usual $\ket{0}$ state conditioned on the previous measurement outcome.

\begin{small}
\begin{align}
\Qcircuit @C=0.7em @R=0.6em {
\lstick{y_1} && \cctrlo{5} & \cctrl{5}&  &\\
\lstick{y_2} && \cw &\cw&\cctrlo{5} & \cctrl{5}&  &\\
\lstick{} & \vdots   & & & &&&\dots &  & \vdots&  & \vdots& & \\
\lstick{y_{n-1}} && \cw &\cw&\cw & \cw& \cw&\dots && \cw &\cw &\cctrlo{5} & \cctrl{5}&  \\ 
\lstick{y_n} & &\cw &\cw&\cw & \cw \cw& \cw &\dots&& \cw & \cw&\cw& \cw  &\cctrlo{5} & \cctrl{5}&\\
\lstick{\ket{0^1}} &&\multigate{5}{Z_{\phi_1}^{y_1\oplus 1}}&  \multigate{5}{X_{\theta_1}^{y_1}}& \qw &  \qw &  \qw & \dots& & \qw & \qw&\qw& \qw & \qw&\qw& \qw &\measureD{Z^{(1)}}\\
\lstick{\ket{0^2}} &&\ghost{Z_{\phi_1}^{y_1\oplus 1}}&  \ghost{X_{\theta_1}^{y_1}}& \multigate{4}{Z_{\phi_2}^{y_2\oplus 1}}&  \multigate{4}{X_{\theta_2}^{y_2}}& \qw & \dots & &\qw& \qw & \qw&\qw & \qw & \qw&\qw &\measureD{Z^{(2)}}\\
\lstick{} &&\ghost{Z_{\phi_1}^{y_1\oplus 1}}&  \ghost{X_{\theta_1}^{y_1}}& \ghost{Z_{\phi_2}^{y_2\oplus 1}}&  \ghost{X_{\theta_2}^{y_2}}& \qw & \dots & & \vdots& & & &  & &\vdots\\
\lstick{\ket{0^{n-1}}} &&\ghost{Z_{\phi_1}^{y_1\oplus 1}}&  \ghost{X_{\theta_1}^{y_1}}& \ghost{Z_{\phi_2}^{y_2\oplus 1}}&  \ghost{X_{\theta_2}^{y_2}}& \qw & \dots &&\qw& \qw  & \multigate{2}{Z_{\phi_{n-1}}^{y_{n-1}\oplus 1}}&\multigate{2}{X_{\theta_{n-1}}^{y_{n-1}}} \qw&\qw& \qw &\qw& \measureD{Z^{(n-1)}}\\
\lstick{\ket{0^n}} &&\ghost{Z_{\phi_1}^{y_1\oplus 1}}&  \ghost{X_{\theta_1}^{y_1}}& \ghost{Z_{\phi_2}^{y_2\oplus 1}}&  \ghost{X_{\theta_2}^{y_2}}& \qw & \dots & &\qw& \qw &\ghost{Z_{\phi_{n-1}}^{y_{n-1}\oplus 1}}&\ghost{X_{\theta_{n-1}}^{y_{n-1}}}&\multigate{1}{Z_{0}^{y_n\oplus 1}} & \multigate{1}{X_{0}^{y_n}}& \qw&\measureD{Z^{(n)}}\\
\lstick{\rho_B}&&\ghost{Z_{\phi_1}^{y_1\oplus 1}} & \ghost{X_{\theta_1}^{y_1}}& \ghost{Z_{\phi_2}^{y_2\oplus 1}}&  \ghost{X_{\theta_2}^{y_2}}& \qw & \dots & &\qw& \qw & \ghost{Z_{\phi_{n-1}}^{y_{n-1}\oplus 1}}&\ghost{X_{\theta_{n-1}}^{y_{n-1}}}&\ghost{Z_{0}^{y_n\oplus 1}} & \ghost{X_{0}^{y_n}}& \qw
} \label{circuitclassicalcontrol}
\end{align}
\end{small}

\section{Numerical Results} \label{sec:numericalresults}

\subsection{Ideal Case} \label{ssec:ideal}

\begin{figure}[h!]
  \begin{subfigure}[t]{0.3\textwidth}
        \centering
            \includegraphics[width = 1.3\linewidth, height= 0.9\linewidth]{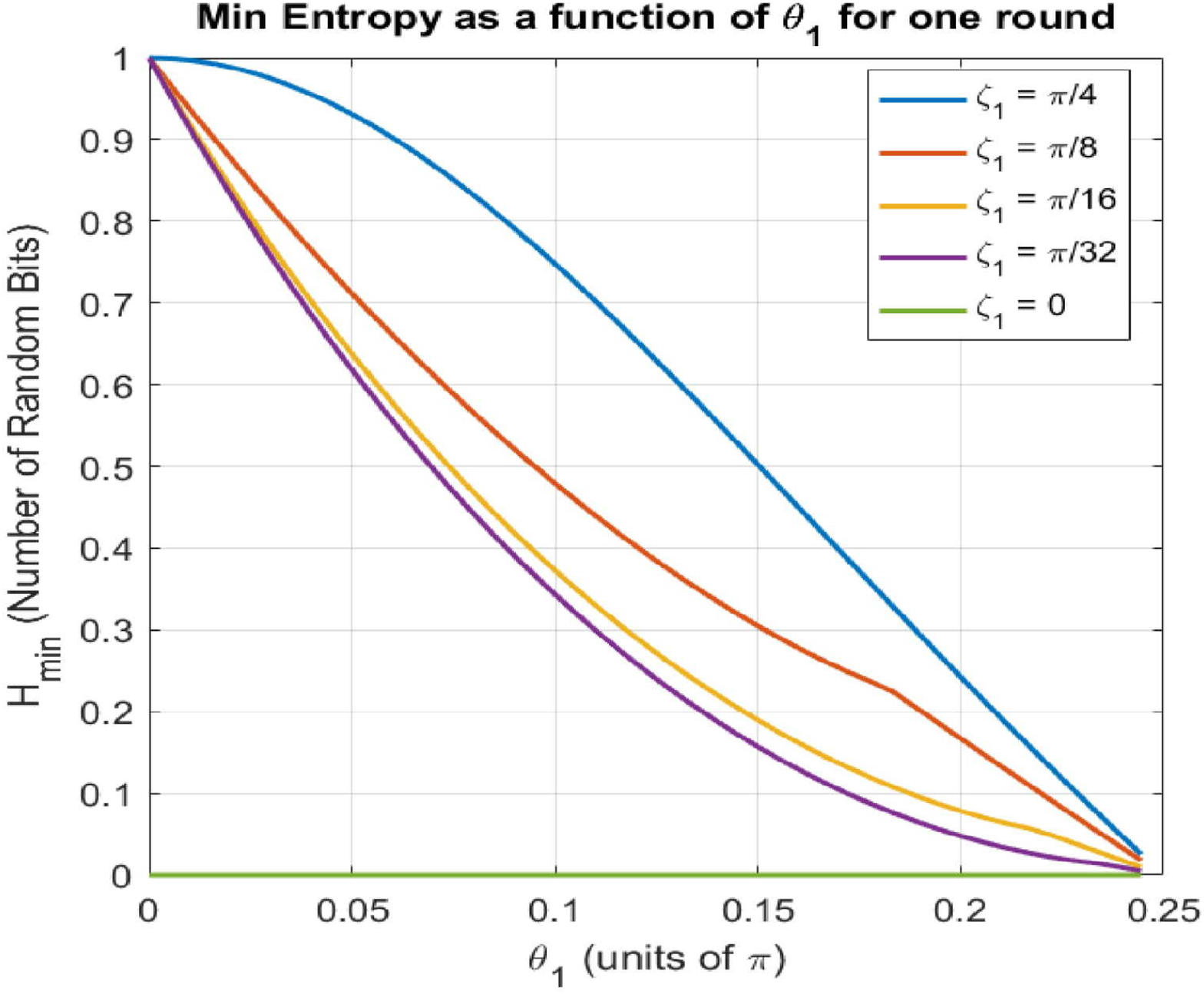}
            \caption{One round of measurements, for a range of initial states, $\zeta_1$, as a function of initial measurement angle, $\theta_1$}
            \label{fig:OneRound}
    \end{subfigure}%
\qquad
    \begin{subfigure}[t]{0.3\textwidth}
        \centering
        \includegraphics[width = 1.3\linewidth, height= 0.9\linewidth]{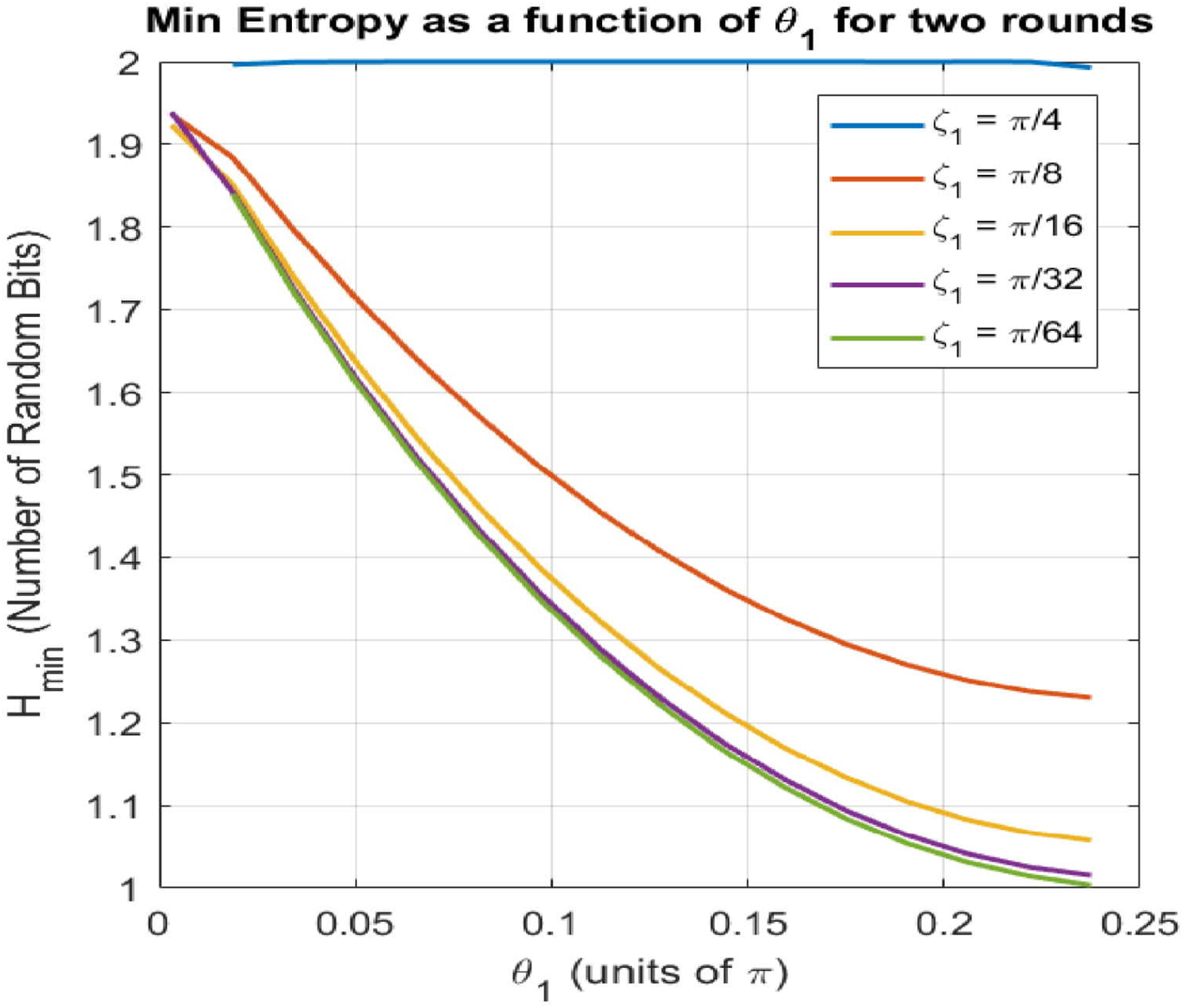}
                \caption{$H_{min}$ for two rounds of measurements, with a range of initial states, $\zeta_1 \in (0,\frac{\pi}{4}]$ and $\phi_1 = \theta_2 = \phi_2 =0$. \label{fig:TwoRounds1}}
    \end{subfigure}%
\qquad
    \begin{subfigure}[t]{0.3\textwidth}
        \includegraphics[width = 1.3\linewidth, height= 0.9\linewidth]{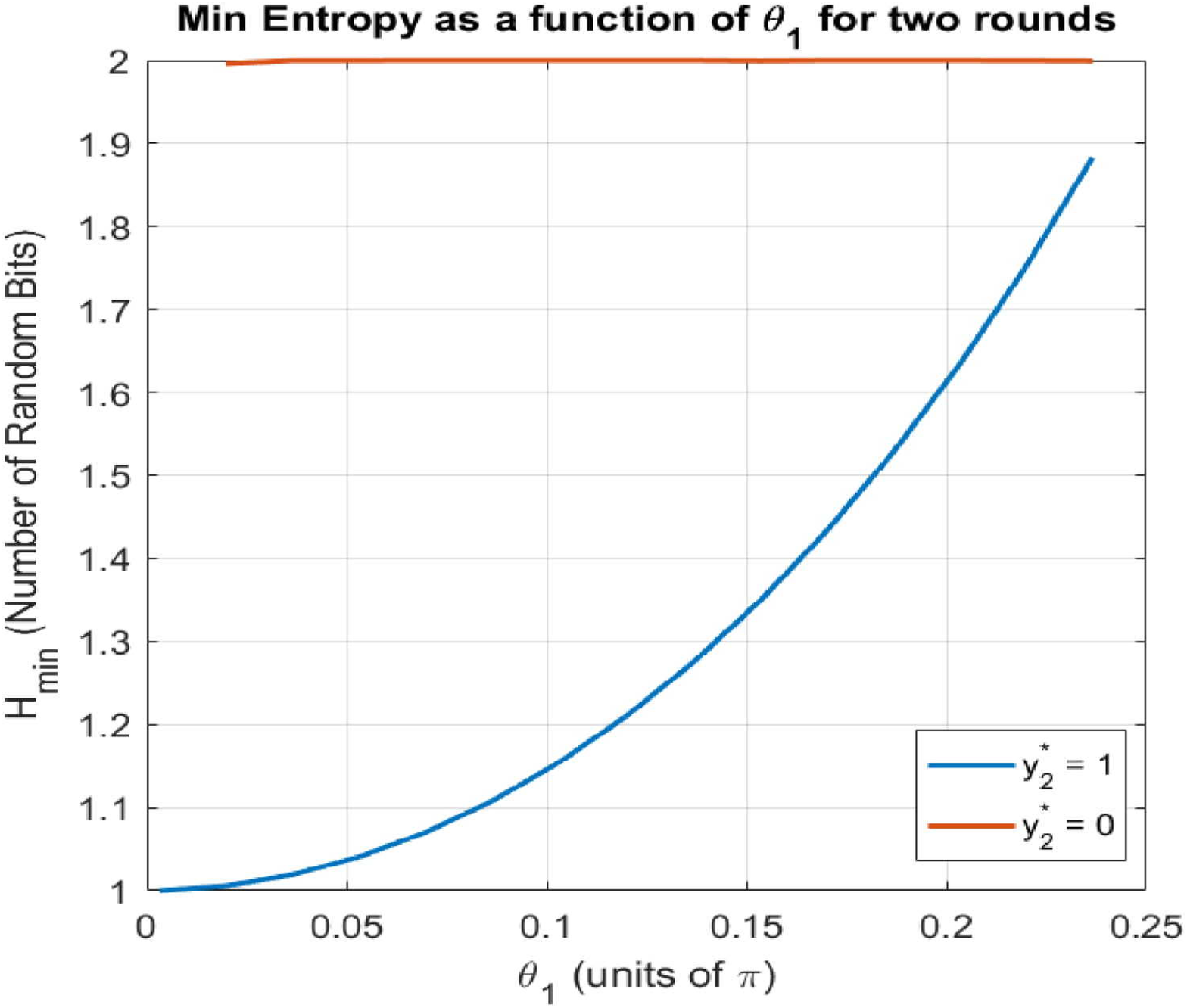}
        \caption{Difference in certified randomness when choosing between measurement settings $y^*_2 = 1 \text{ or } y^*_2  = 0$ in the second measurement round.\label{fig:TwoRounds2}}
    \end{subfigure}
\caption{Min. Entropy, $H_{min}$, for one and two measurement rounds, as a function of initial measurement angle, $\theta_1$.} \label{fig:OneTwoRounds}
\end{figure}

In this section, numerical results are presented to illustrate the performance of the protocol, in particular the SDP, (\ref{mainsdp}), assuming ideal functionality of both devices. All numerical results were obtained using the Matlab convex optimization package, \textit{cvx}, \cite{boyd_convex_2004} and a package for managing quantum states, \textit{qetlab}, \cite{johnston_qetlab:_2016}. The codes can be found at \cite{briancoyle_tpmscproject2017:_2017}. As a convention, it will be assumed that Bob always measures in the noisy X basis in the first round, with the final measurement round in the protocol being projective, $\theta_n = 0$ or $\phi_n = 0$, depending on whether $n$ is odd or even. In all of the following, we assume the measurement statistics and runs of the protocol are i.i.d.\\
\indent For completeness, the Min. Entropy for one round of measurement is plotted as a function of measurement angles used for the first round, with the noisy X measurements, $X_{\theta_1}$, for a range of values of $\theta_1$, as seen in Figure~(\ref{fig:OneRound}). All measurements are applied on the initial state:
\begin{align}
    \ket{\Psi(\zeta_1)} &= \cos(\zeta_1)\ket{00}+\sin(\zeta_1)\ket{11} \label{initialstate}
\end{align}
$\ket{\Psi(\zeta_1)}$ was measured for values of: $\zeta_1 \in \{0, \frac{\pi}{32}, \frac{\pi}{16}, \frac{\pi}{8}, \frac{\pi}{4} \}$. The solution of this SDP clearly reproduce the already known results for a single measurement round, as is done in \cite{skrzypczyk_steeringreview:_2016, passaro_optimal_2015}, but using our SDP which is slightly different than the one derived in that paper and requires no assumption about the specific actions of the adversary, Eve. As expected, when $\zeta_1 = 0$, no randomness can be certified as the state becomes a product state, whereas for $\zeta_1 = \pi/4$, the maximal amount of randomness can be certified, since this state is maximally entangled between Alice and Bob.

\begin{figure}[h!]
    \begin{subfigure}[t]{0.3\linewidth}
        \centering
        \includegraphics[width = 1.5\linewidth, height= 0.9\linewidth]{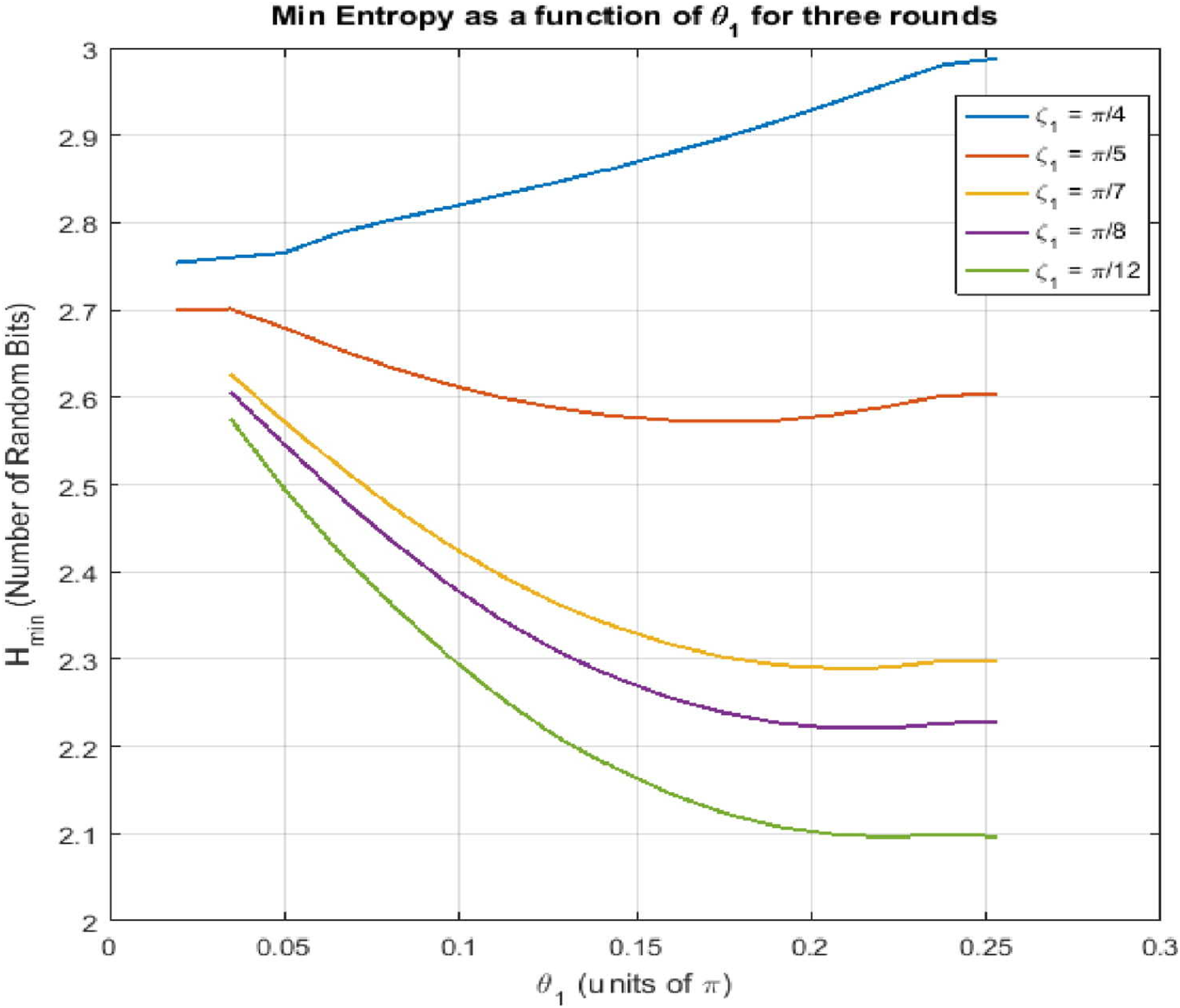}
        \caption{$H_{min}$ using various initial states, with initial angles, $\zeta_1 \in \{\frac{\pi}{4}, \frac{\pi}{5}, \frac{\pi}{7}, \frac{\pi}{8}, \frac{\pi}{12}\}$.}
        \label{fig:ThreeRounds1}
        \end{subfigure}%
        \qquad
        \begin{subfigure}[t]{0.3\linewidth}
            \centering
             \includegraphics[width =1.5\linewidth, height = 0.9\linewidth]{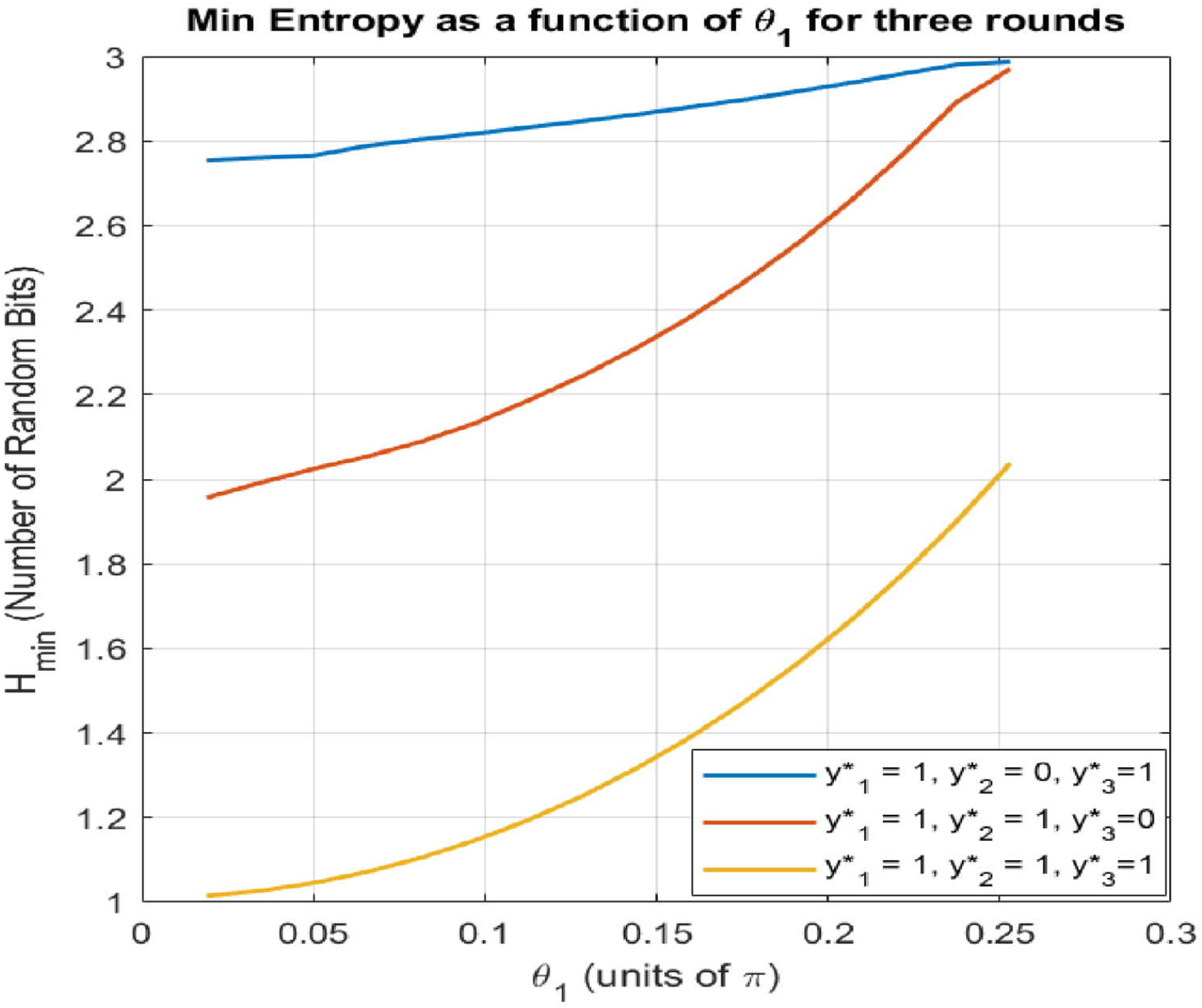}
             \caption{$H_{min}$ using various measurement settings, $y_1^*, y_2^*, y_3^*$.}
             \label{fig:ThreeRounds2}
        \end{subfigure}%
        \qquad
        \begin{subfigure}[t]{0.3\linewidth}
            \centering
             \includegraphics[width =1.5\linewidth, height =0.9\linewidth]{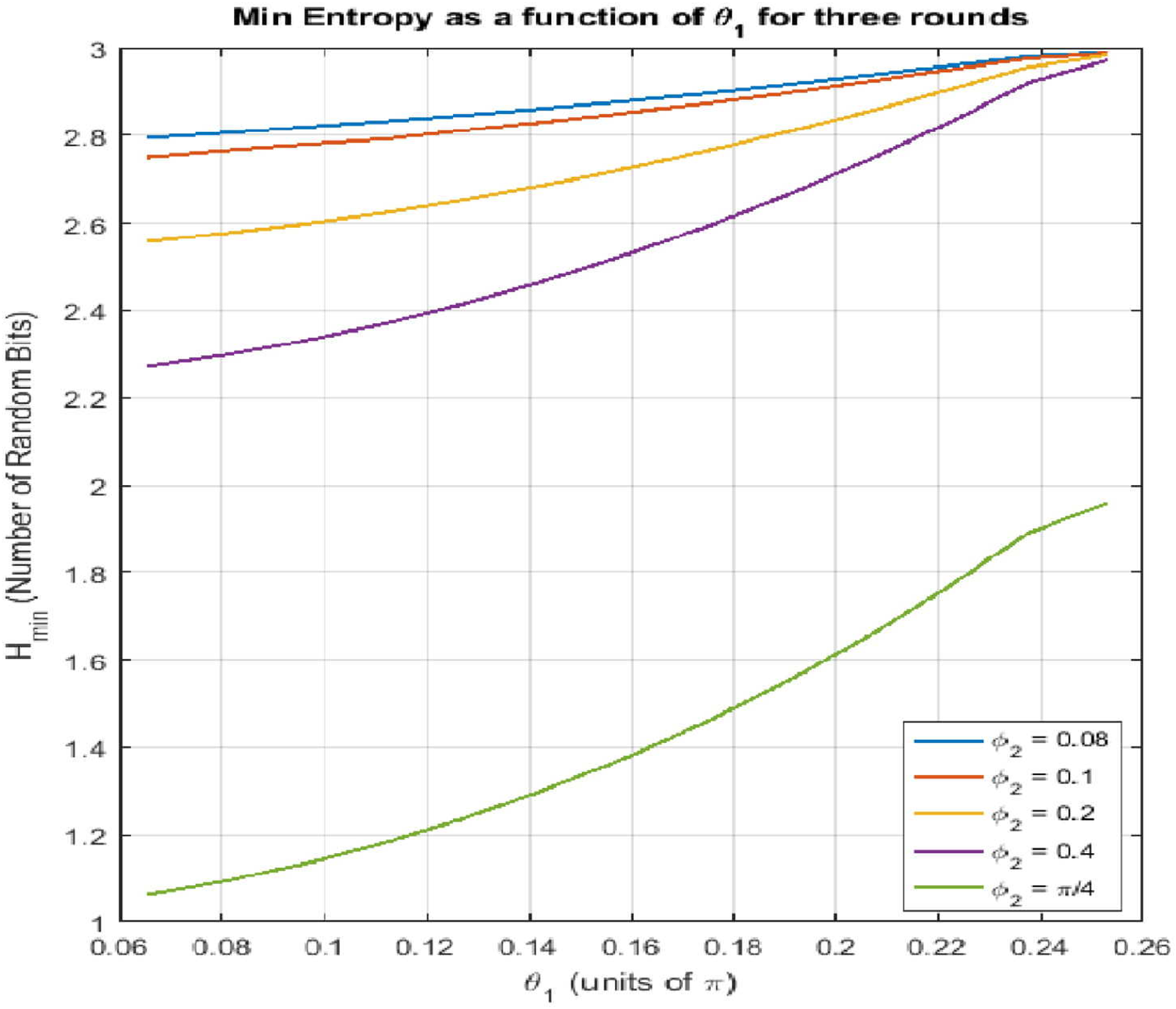}
             \caption{$H_{min}$ using various angles in the second round, $\phi_2 \in \{0.08 , 0.1 , 0.2, 0.4, \frac{\pi}{4}\}$ rad.}
             \label{fig:ThreeRounds3}
        \end{subfigure}
    \caption{$H_{min}$ for three measurement rounds, as a function of initial measurement angle, $\theta_1$.} \label{fig:ThreeRounds}
\end{figure}

Figures~(\ref{fig:TwoRounds1}) and (\ref{fig:TwoRounds2}) show the results after two measurement rounds. In Figure~(\ref{fig:TwoRounds1}), the measurement in round one was taken to be in the noisy $X$ basis, with a range of initial angles $\zeta_1$, and the measurement in round two was taken to be in the usual computational basis, $\phi_2 = 0$. Figure~(\ref{fig:TwoRounds2}) illustrates the difference in choosing different measurement choices for the second round, i.e. between $y_2^* = 0$, or $y_2^* = 1$, with maximal randomness certified after sequential measurements in alternating bases, for example $y_1^* = 1, y_2^* = 0$. \\
\indent Finally, Figure~(\ref{fig:ThreeRounds}) shows numerical results for the protocol for three measurement rounds. The protocol proceeds in exactly the same manner as for one and two rounds. In particular, in the first round, Bob can choose between a non-projective measurement in the noisy $X_{\theta_1}$ basis, or if the particular run of the protocol is a test, he will measure in the projective $Z^{0}$ basis. In the second round, he will choose to measure in the noisy $Z_{\phi_2}$ basis, or the $X_{0}$ basis for a test run. In the final round, he will choose to measure in the projective ($\theta_3 = 0$) $X_{0}$ basis, or the  projective ($\phi_3 = 0$) $Z_{0}$ basis for a test. Again, Figure~(\ref{fig:ThreeRounds2}) reiterates the optimality of using an alternating sequence of non-projective measurements, with the most randomness produced with the setting $y_1^* = 1, y_2^* = 0, y_3^* = 1$ in this example. Figure~(\ref{fig:ThreeRounds3}) shows the results for various second round measurement angles, and the amount of randomness that can be certified increases as the measurement angle, $\phi_2\rightarrow 0$.

\subsection{Networked Ion Trap Implementation}\label{ssec:protocolimplementation}

The framework in which we have designed this protocol, assuming a \textit{malicious} adversary, Eve, is general enough to include the scenario in which she is not intentionally trying to interfere with our randomness generation, but instead we can imagine that Eve simply made some error in building the devices. This would correspond to introducing some noise, for example, in our state preparation and/or measurement apparatus. This noise assumption is clearly a sub-case of the malicious adversary scenario. This mentality allows us to use our protocol to evaluate the usefulness of some current available technologies for randomness generation purposes, in some simple cases. In particular, we will restrict to assuming we only have some noise in our state preparation, but all other parts of the device works perfectly. To do so, we test the state introduced in \cite{nigmatullin_minimally_2016}, which can be produced between two parties in a networked architecture of ion traps:
\begin{align}
    \rho^{(0)}_\epsilon = (1-\epsilon)\Phi^+ + \epsilon/3\Phi^-+\epsilon/3\Psi^+ +\epsilon/3\Psi^- \label{rawstate1}
\end{align}
where $\phi^+, \phi^-, \psi^+, \psi^-$ are the standard 2-qubit Bell states. The state, (\ref{rawstate1}), is  a mixed state assuming uniform depolarising noise. In \cite{nigmatullin_minimally_2016}, this  state is assumed to be one produced by two ion traps entangled by a photonic link. The simple noise model is chosen to allow use of a technique to purify the state. In particular, after 3 rounds of this purification protocol, the resulting states are given by:

\begin{align}
    \rho^{(1)}_\epsilon &= \left(1-2/3\epsilon-2/3\epsilon^2\right)\Phi^+ +\left(2/9\epsilon+2/9\epsilon^2\right)\Phi^-  + 2/9\epsilon^2\Psi^+ + 2/9\epsilon^2\Psi^- + O(\epsilon^3) \label{1roundpure}\\ 
    \rho^{(2)}_\epsilon &= \left(1- 8/9\epsilon^2- 8/27 \epsilon^3\right)\Phi^+ +  4/9\epsilon^2\Phi^-
    + 4/9\epsilon^2\Psi^+ + 8/27\epsilon^3\Psi^- + O(\epsilon^4)
    \label{tworoundpure}\\ 
    \rho^{(3)}_\epsilon &= \left(1- 2/9\epsilon^2- 16/27\epsilon^3\right)\Phi^+ + 2/9\epsilon^2\Phi^- + 8/27\epsilon^3\Psi^+ + 8/27\epsilon^3\Psi^- + O(\epsilon^4) \label{threeroundpure}
\end{align}

where $\rho^{(i)}_\epsilon$ is the state produced after $i$ rounds of the purification protocol. \\
\indent Currently, raw entanglement between two ion traps, connected with an entangling photon, has been achieved with a fidelity of about $85\% \implies \epsilon \approx 0.15$, \cite{hucul_modular_2014}. Starting with this level of raw infidelity, the purification protocol produces states of infidelity $ \epsilon \approx 0.1, 0.02, 0.005$ after one, two and three rounds respectively. The fidelity is given by (\ref{fidelity}), \cite{nielsen_quantum_2011}, and taken to be between the actual state $\rho^{(i)}$, and the pure Bell state, $\Phi^+$:
\begin{align}
    F(\rho^{(i)}_{\epsilon},\Phi^+) = \Tr\left(\sqrt{\sqrt{\rho^{(i)}_\epsilon}\Phi^+\sqrt{\rho^{(i)}_\epsilon}}\right)  \label{fidelity}
\end{align}

\begin{figure}[h]
    \begin{subfigure}[t]{0.45\linewidth}
        \centering
        \includegraphics[width =1.3\linewidth, height =0.6\linewidth]{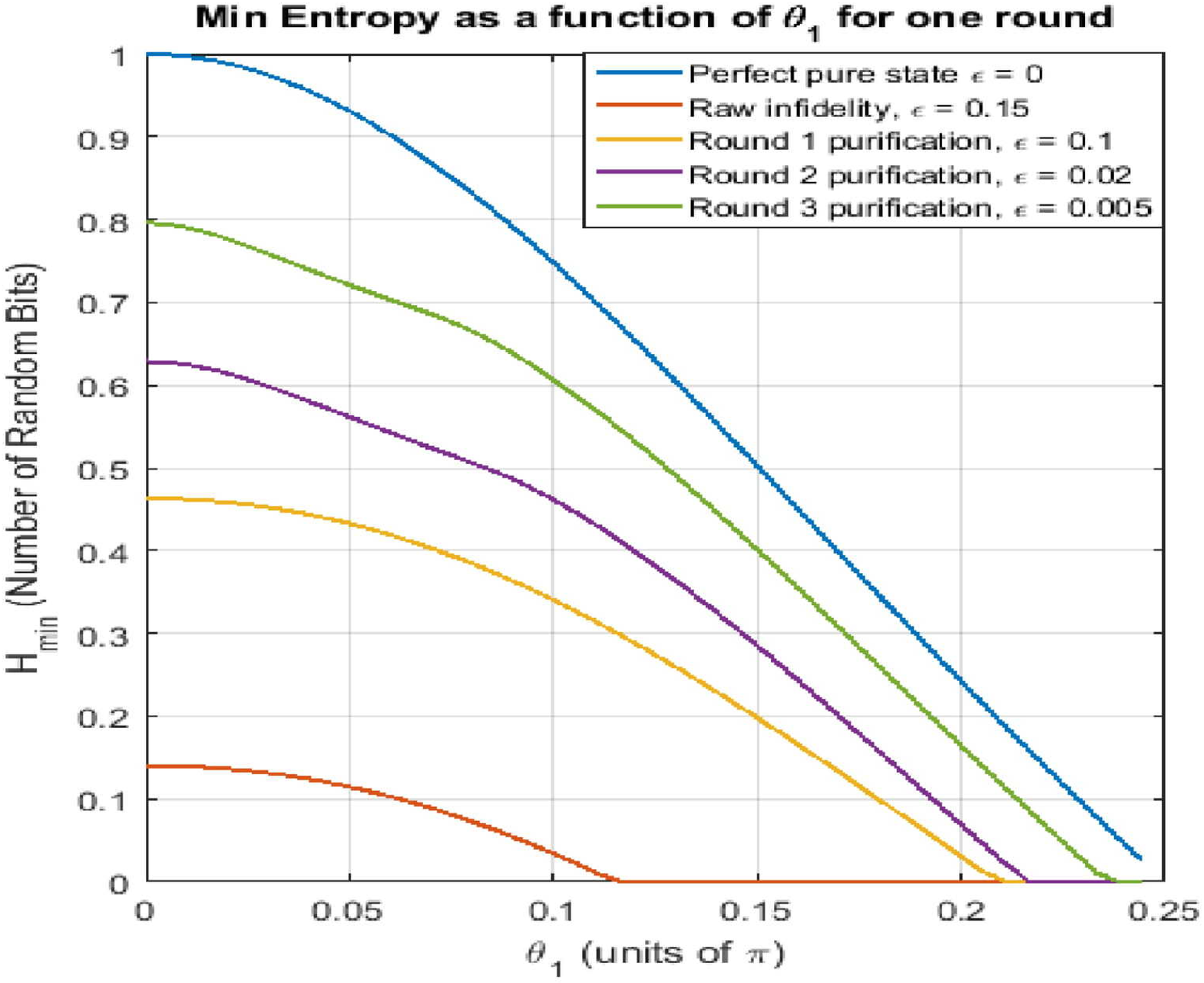}
        \caption{Single measurement on the raw entangled state (\ref{rawstate1}) ($\epsilon = 0.15$), the states produced after three rounds of the purification protocol, (\ref{1roundpure}, \ref{tworoundpure}, \ref{threeroundpure}), with $\epsilon = 0.1, 0.02, 0.005$ respectively and a perfect pure state with $\epsilon =0 $.}
        \label{fig:oneroundpure}
    \end{subfigure}%
\qquad
    \begin{subfigure}[t]{0.45\linewidth}
        \centering
        \includegraphics[width =1.3\linewidth, height =0.6\linewidth]{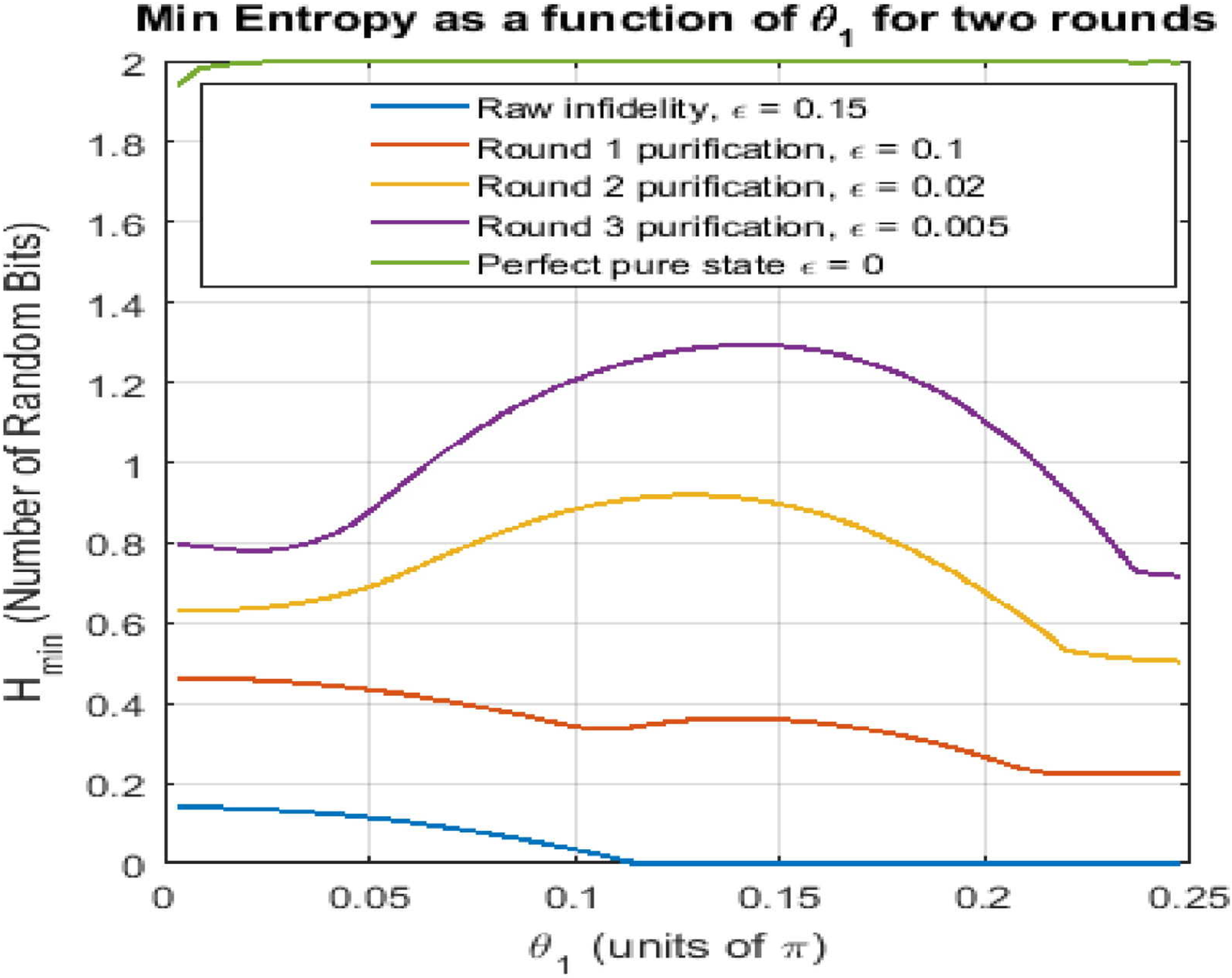}
        \caption{Two rounds of measurement on the raw entangled state (\ref{rawstate1}) ($\epsilon = 0.15$), the states produced after three rounds of the purification protocol, (\ref{1roundpure}, \ref{tworoundpure}, \ref{threeroundpure}), with $\epsilon = 0.1, 0.02, 0.005$ respectively and a perfect pure state with $\epsilon =0 $.}
        \label{fig:tworoundspure}
    \end{subfigure}
    \caption{1SDI protocol implemented for one and two measurement rounds on noisy states produced by a networked ion-trap architecture.}
\end{figure}

Given the levels of entanglement present in the states above, we test the advantage of using a sequence of measurements vs. a single measurement on a noisy entangled state. Figure~(\ref{fig:oneroundpure}) shows the result after a single X measurement on the states (choosing $y^*_1 = 1$) (\ref{rawstate1}, \ref{1roundpure}, \ref{tworoundpure}, \ref{threeroundpure}). Clearly, maximal randomness can be certified in the case where the measurement is projective, as expected. It can also be seen that by using the raw entangled state, (\ref{rawstate1}), very little randomness can be certified, with a maximum of approximately 0.15 bits.\\
\indent Figure~(\ref{fig:tworoundspure}) illustrates the results after two rounds of measurements, where the second round measurements are projective, $\theta_2 = \phi_2 = 0$. The case of $\theta_1 = 0$ gives the same result as the single measurement scenario, since in this case the first measurement is projective and hence no randomness can be certified in the second round.\\
\indent Unfortunately, it can be seen that no extra randomness can be certified in two measurement rounds on the raw entangled state, (\ref{rawstate1}). However, after two or more rounds of the purification protocol, indeed more randomness can be certified by using a sequence vs. a single measurement, as indicated by the peaks in Figure~(\ref{fig:tworoundspure}). The infidelity for which the sequence becomes more useful than a single measurement can be seen to be approximately in the interval $\epsilon \in (0.06, 0.07)$.

\begin{figure}[h!]
\begin{subfigure}[t]{0.45\linewidth}
\centering
        \includegraphics[width =1.3\linewidth, height =0.8\linewidth]{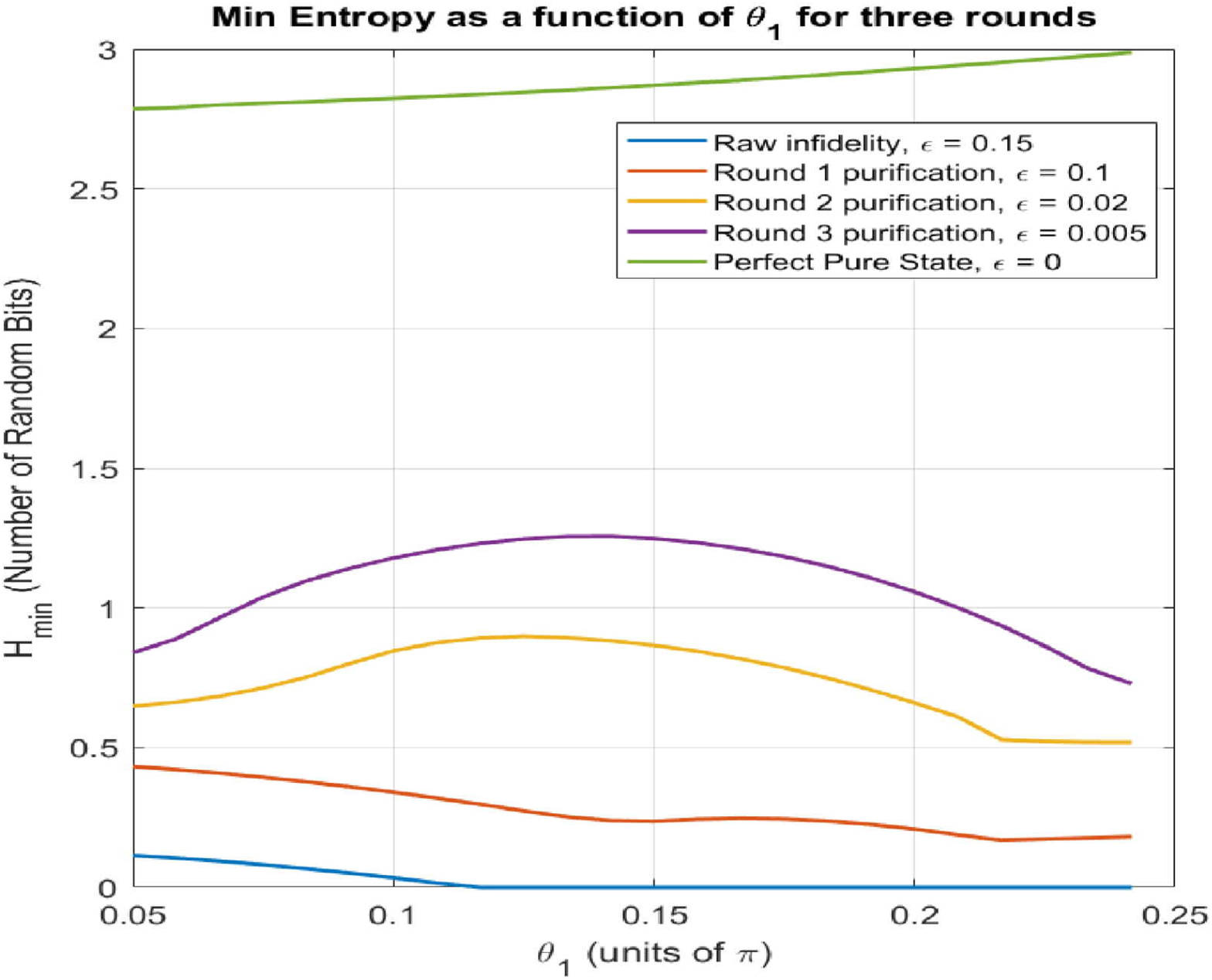}
        \caption{Three rounds of measurement on the raw entangled state, (\ref{rawstate1}), ($\epsilon = 0.15$), the states produced after three rounds of the purification protocol, (\ref{1roundpure}, \ref{tworoundpure}, \ref{threeroundpure}), with $\epsilon = 0.1, 0.02, 0.005$ respectively and a perfect pure state with $\epsilon =0 $.}
        \label{fig:threeroundspure1}
    \end{subfigure}
\qquad
    \begin{subfigure}[t]{0.45\linewidth}
        \centering
        \includegraphics[width =1.3\linewidth, height =0.8\linewidth]{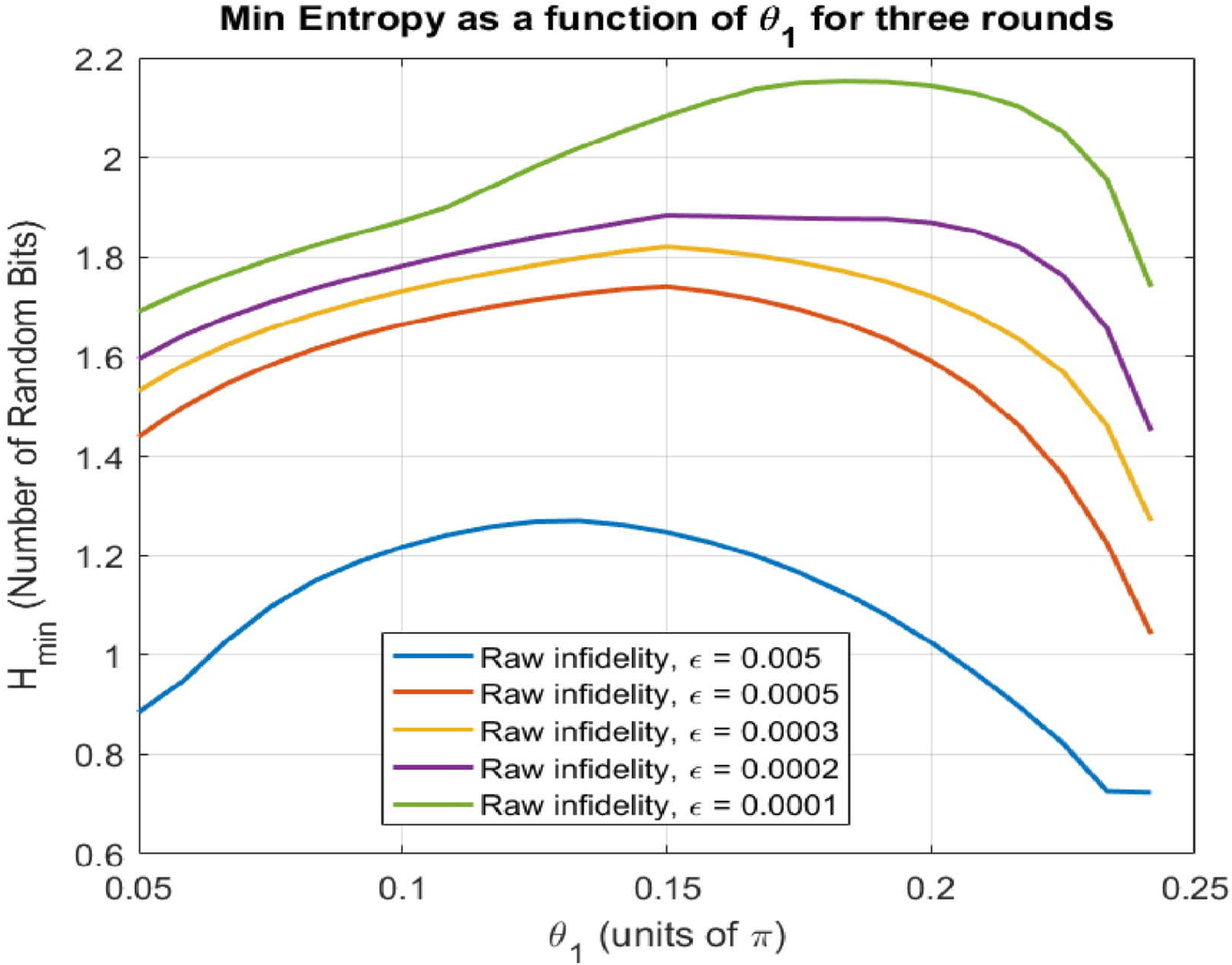}
        \caption{Three rounds of measurement on raw entangled states with infidelities $\epsilon = \{5\times 10^{-3}, 5\times 10^{-4}, 3\times 10^{-4}, 2\times 10^{-4},  1\times 10^{-4}\}$.\label{fig:threeroundspure2}}
    \end{subfigure}
    \caption{Three rounds of measurements on states with various levels of entanglement fidelity.}
    \label{fig:threeroundspure}
\end{figure}

Finally, Figure~(\ref{fig:threeroundspure1}) shows the results after three rounds of measurements, where the third, and final round of measurements are projective with $\theta_3 = \phi_3 = 0$. The second round of measurements is chosen in this case to be a noisy $Z$ measurement, with $\phi_2 = 0.08 \text{ rad}$.\\
\indent Unfortunately, it can be seen that no extra randomness can be certified by implementing three measurements, than with two rounds. This is even the case for the purified states, (\ref{1roundpure}, \ref{tworoundpure}, \ref{threeroundpure}), so even these levels of purity are not sufficient to extract more randomness from a single state with three rounds of measurements. The perfect pure state, with $\epsilon = 0$ is also plotted for comparison.\\
\indent Clearly, there must be some level at which the state becomes pure enough to be useful so Figure~(\ref{fig:threeroundspure2}) shows the results of the protocol for very small infidelities, specifically:
\begin{align*}
\epsilon = \{5\times 10^{-3}, 5\times 10^{-4}, 3\times 10^{-4}, 2\times 10^{-4}, 1\times 10^{-4}\}
\end{align*}
It can be seen that for an infidelity approximately in the interval,  $\epsilon \in (1\times 10^{-4}, 2\times 10^{-4})$, the state is pure enough to be able to certify more randomness with three rounds of measurement, than with two. This corresponds to being able to create pure entangled states experimentally with fidelities of greater than $99.98 \%$. This level could be reached by repeating the purification protocol more times but clearly this decreases the efficiency of the protocol as many more extra qubits would need to be introduced to implement this purification. It is expected that for 4 and higher rounds of measurement, states which have an even higher level of purity would be required to make the protocol worthwhile, i.e. so that rounds of measurements on a single state would give better results than single measurements on new states each time.

\bibliographystyle{eptcs}
\bibliography{Paper}

\begin{thebibliography}{10}
\providecommand{\bibitemdeclare}[2]{}
\providecommand{\surnamestart}{}
\providecommand{\surnameend}{}
\providecommand{\urlprefix}{Available at }
\providecommand{\url}[1]{\texttt{#1}}
\providecommand{\href}[2]{\texttt{#2}}
\providecommand{\urlalt}[2]{\href{#1}{#2}}
\providecommand{\doi}[1]{doi:\urlalt{http://dx.doi.org/#1}{#1}}
\providecommand{\bibinfo}[2]{#2}

\bibitemdeclare{article}{bell_einstein_1964}
\bibitem{bell_einstein_1964}
\bibinfo{author}{J.~S. \surnamestart Bell\surnameend} (\bibinfo{year}{1964}):
  \emph{\bibinfo{title}{On the {Einstein} {Podolsky} {Rosen} paradox}}.
\newblock {\sl \bibinfo{journal}{Physics Physique Fizika}}
  \bibinfo{volume}{1}(\bibinfo{number}{3}), pp. \bibinfo{pages}{195--200},
  \doi{10.1103/PhysicsPhysiqueFizika.1.195}.
\newblock
  \urlprefix\url{https://link.aps.org/doi/10.1103/PhysicsPhysiqueFizika.1.195}.

\bibitemdeclare{misc}{boyd_convex_2004}
\bibitem{boyd_convex_2004}
\bibinfo{author}{Stephen \surnamestart Boyd\surnameend} \&
  \bibinfo{author}{Lieven \surnamestart Vandenberghe\surnameend}
  (\bibinfo{year}{2004}): \emph{\bibinfo{title}{Convex {Optimization} by
  {Stephen} {Boyd}}}, \doi{10.1017/CBO9780511804441}.
\newblock
  \urlprefix\url{/core/books/convex-optimization/17D2FAA54F641A2F62C7CCD01DFA97C4}.

\bibitemdeclare{book}{briancoyle_tpmscproject2017:_2017}
\bibitem{briancoyle_tpmscproject2017:_2017}
\bibinfo{author}{\surnamestart {BrianCoyle}\surnameend} (\bibinfo{year}{2017}):
  \emph{\bibinfo{title}{{TPMScProject}2017: {MSc} {Project} {Codes} {For} 1SDI
  {Certification} of {Random} {Numbers}}}.
\newblock \doi{10.5281/zenodo.1257182}.
\newblock \urlprefix\url{https://github.com/BrianCoyle/TPMScProject2017}.

\bibitemdeclare{article}{colbeck_quantum_2009}
\bibitem{colbeck_quantum_2009}
\bibinfo{author}{Roger \surnamestart Colbeck\surnameend}
  (\bibinfo{year}{2009}): \emph{\bibinfo{title}{Quantum {And} {Relativistic}
  {Protocols} {For} {Secure} {Multi}-{Party} {Computation}}}.
\newblock {\sl \bibinfo{journal}{arXiv:0911.3814 [quant-ph]}}.
\newblock \urlprefix\url{http://arxiv.org/abs/0911.3814}.
\newblock \bibinfo{note}{ArXiv: 0911.3814}.

\bibitemdeclare{article}{curchod_unbounded_2017}
\bibitem{curchod_unbounded_2017}
\bibinfo{author}{F.~J. \surnamestart Curchod\surnameend},
  \bibinfo{author}{M.~\surnamestart Johansson\surnameend},
  \bibinfo{author}{R.~\surnamestart Augusiak\surnameend},
  \bibinfo{author}{M.~J. \surnamestart Hoban\surnameend},
  \bibinfo{author}{P.~\surnamestart Wittek\surnameend} \&
  \bibinfo{author}{A.~\surnamestart Acín\surnameend} (\bibinfo{year}{2017}):
  \emph{\bibinfo{title}{Unbounded randomness certification using sequences of
  measurements}}.
\newblock {\sl \bibinfo{journal}{Phys. Rev. A}}
  \bibinfo{volume}{95}(\bibinfo{number}{2}), p. \bibinfo{pages}{020102},
  \doi{10.1103/PhysRevA.95.020102}.
\newblock \urlprefix\url{https://link.aps.org/doi/10.1103/PhysRevA.95.020102}.

\bibitemdeclare{article}{einstein_can_1935}
\bibitem{einstein_can_1935}
\bibinfo{author}{A.~\surnamestart Einstein\surnameend},
  \bibinfo{author}{B.~\surnamestart Podolsky\surnameend} \&
  \bibinfo{author}{N.~\surnamestart Rosen\surnameend} (\bibinfo{year}{1935}):
  \emph{\bibinfo{title}{Can {Quantum}-{Mechanical} {Description} of {Physical}
  {Reality} {Be} {Considered} {Complete}?}}
\newblock {\sl \bibinfo{journal}{Phys. Rev.}}
  \bibinfo{volume}{47}(\bibinfo{number}{10}), pp. \bibinfo{pages}{777--780},
  \doi{10.1103/PhysRev.47.777}.
\newblock \urlprefix\url{https://link.aps.org/doi/10.1103/PhysRev.47.777}.

\bibitemdeclare{article}{hucul_modular_2014}
\bibitem{hucul_modular_2014}
\bibinfo{author}{D.~\surnamestart Hucul\surnameend}, \bibinfo{author}{I.~V.
  \surnamestart Inlek\surnameend}, \bibinfo{author}{G.~\surnamestart
  Vittorini\surnameend}, \bibinfo{author}{C.~\surnamestart Crocker\surnameend},
  \bibinfo{author}{S.~\surnamestart Debnath\surnameend}, \bibinfo{author}{S.~M.
  \surnamestart Clark\surnameend} \& \bibinfo{author}{C.~\surnamestart
  Monroe\surnameend} (\bibinfo{year}{2014}): \emph{\bibinfo{title}{Modular
  entanglement of atomic qubits using photons and phonons}}.
\newblock {\sl \bibinfo{journal}{Nature Physics}} \bibinfo{volume}{11},
  p.~\bibinfo{pages}{37}.
\newblock \urlprefix\url{http://dx.doi.org/10.1038/nphys3150}.

\bibitemdeclare{misc}{johnston_qetlab:_2016}
\bibitem{johnston_qetlab:_2016}
\bibinfo{author}{Nathaniel \surnamestart Johnston\surnameend},
  \bibinfo{author}{Alessandro \surnamestart Cosentino\surnameend} \&
  \bibinfo{author}{Vincent \surnamestart Russo\surnameend}
  (\bibinfo{year}{2016}): \emph{\bibinfo{title}{{QETLAB}: {A} {MATLAB} toolbox
  for quantum entanglement, version 0.9}}, \doi{10.5281/zenodo.44637}.
\newblock \urlprefix\url{http://qetlab.com}.

\bibitemdeclare{article}{law_quantum_2014}
\bibitem{law_quantum_2014}
\bibinfo{author}{Y.~Z. \surnamestart Law\surnameend}, \bibinfo{author}{L.~P.
  \surnamestart Thinh\surnameend}, \bibinfo{author}{J.-D. \surnamestart
  Bancal\surnameend} \& \bibinfo{author}{V.~\surnamestart Scarani\surnameend}
  (\bibinfo{year}{2014}): \emph{\bibinfo{title}{Quantum randomness extraction
  for various levels of characterization of the devices}}.
\newblock {\sl \bibinfo{journal}{J. Phys. A: Math. Theor.}}
  \bibinfo{volume}{47}(\bibinfo{number}{42}), p. \bibinfo{pages}{424028},
  \doi{10.1088/1751-8113/47/42/424028}.
\newblock \urlprefix\url{http://stacks.iop.org/1751-8121/47/i=42/a=424028}.

\bibitemdeclare{book}{nielsen_quantum_2011}
\bibitem{nielsen_quantum_2011}
\bibinfo{author}{Michael~A. \surnamestart Nielsen\surnameend} \&
  \bibinfo{author}{Isaac~L. \surnamestart Chuang\surnameend}
  (\bibinfo{year}{2011}): \emph{\bibinfo{title}{Quantum {Computation} and
  {Quantum} {Information}: 10th {Anniversary} {Edition}}},
  \bibinfo{edition}{10th} edition.
\newblock \bibinfo{publisher}{Cambridge University Press},
  \bibinfo{address}{New York, NY, USA}.

\bibitemdeclare{article}{nigmatullin_minimally_2016}
\bibitem{nigmatullin_minimally_2016}
\bibinfo{author}{Ramil \surnamestart Nigmatullin\surnameend},
  \bibinfo{author}{Christopher~J. \surnamestart Ballance\surnameend},
  \bibinfo{author}{Niel~de \surnamestart Beaudrap\surnameend} \&
  \bibinfo{author}{Simon~C. \surnamestart Benjamin\surnameend}
  (\bibinfo{year}{2016}): \emph{\bibinfo{title}{Minimally complex ion traps as
  modules for quantum communication and computing}}.
\newblock {\sl \bibinfo{journal}{New J. Phys.}}
  \bibinfo{volume}{18}(\bibinfo{number}{10}), p. \bibinfo{pages}{103028},
  \doi{10.1088/1367-2630/18/10/103028}.
\newblock \urlprefix\url{http://stacks.iop.org/1367-2630/18/i=10/a=103028}.

\bibitemdeclare{article}{passaro_optimal_2015}
\bibitem{passaro_optimal_2015}
\bibinfo{author}{Elsa \surnamestart Passaro\surnameend},
  \bibinfo{author}{Daniel \surnamestart Cavalcanti\surnameend},
  \bibinfo{author}{Paul \surnamestart Skrzypczyk\surnameend} \&
  \bibinfo{author}{Antonio \surnamestart Acín\surnameend}
  (\bibinfo{year}{2015}): \emph{\bibinfo{title}{Optimal randomness
  certification in the quantum steering and prepare-and-measure scenarios}}.
\newblock {\sl \bibinfo{journal}{New J. Phys.}}
  \bibinfo{volume}{17}(\bibinfo{number}{11}), p. \bibinfo{pages}{113010},
  \doi{10.1088/1367-2630/17/11/113010}.
\newblock \urlprefix\url{http://stacks.iop.org/1367-2630/17/i=11/a=113010}.

\bibitemdeclare{article}{pironio_random_2010}
\bibitem{pironio_random_2010}
\bibinfo{author}{S.~\surnamestart Pironio\surnameend},
  \bibinfo{author}{A.~\surnamestart Acín\surnameend},
  \bibinfo{author}{S.~\surnamestart Massar\surnameend},
  \bibinfo{author}{A.~Boyer de~la \surnamestart Giroday\surnameend},
  \bibinfo{author}{D.~N. \surnamestart Matsukevich\surnameend},
  \bibinfo{author}{P.~\surnamestart Maunz\surnameend},
  \bibinfo{author}{S.~\surnamestart Olmschenk\surnameend},
  \bibinfo{author}{D.~\surnamestart Hayes\surnameend},
  \bibinfo{author}{L.~\surnamestart Luo\surnameend}, \bibinfo{author}{T.~A.
  \surnamestart Manning\surnameend} \& \bibinfo{author}{C.~\surnamestart
  Monroe\surnameend} (\bibinfo{year}{2010}): \emph{\bibinfo{title}{Random
  numbers certified by {Bell}’s theorem}}.
\newblock {\sl \bibinfo{journal}{Nature}}
  \bibinfo{volume}{464}(\bibinfo{number}{7291}), p. \bibinfo{pages}{1021},
  \doi{10.1038/nature09008}.
\newblock \urlprefix\url{https://www.nature.com/articles/nature09008}.

\bibitemdeclare{book}{skrzypczyk_steeringreview:_2016}
\bibitem{skrzypczyk_steeringreview:_2016}
\bibinfo{author}{Paul \surnamestart Skrzypczyk\surnameend}
  (\bibinfo{year}{2016}): \emph{\bibinfo{title}{steeringreview: {Code} to
  accompany "{Quantum} steering: a short review with focus on semi-definite
  programming"}}.
\newblock \urlprefix\url{https://github.com/paulskrzypczyk/steeringreview}.

\bibitemdeclare{article}{skrzypczyk_quantifying_2014}
\bibitem{skrzypczyk_quantifying_2014}
\bibinfo{author}{Paul \surnamestart Skrzypczyk\surnameend},
  \bibinfo{author}{Miguel \surnamestart Navascués\surnameend} \&
  \bibinfo{author}{Daniel \surnamestart Cavalcanti\surnameend}
  (\bibinfo{year}{2014}): \emph{\bibinfo{title}{Quantifying
  {Einstein}-{Podolsky}-{Rosen} {Steering}}}.
\newblock {\sl \bibinfo{journal}{Phys. Rev. Lett.}}
  \bibinfo{volume}{112}(\bibinfo{number}{18}), p. \bibinfo{pages}{180404},
  \doi{10.1103/PhysRevLett.112.180404}.
\newblock
  \urlprefix\url{https://link.aps.org/doi/10.1103/PhysRevLett.112.180404}.

\end{thebibliography}
\end{document}